\documentclass[prb, amsfonts, amssymb, amsmath, reprint, showkeys, nofootinbib, twoside, superscriptaddress,footinbib,floatfix]{revtex4-1}

\usepackage[english]{babel}
\usepackage[utf8]{inputenc}
\usepackage{upgreek}

\usepackage[colorlinks,citecolor=blue,urlcolor=blue,bookmarks=false,hypertexnames=true]{hyperref} 
\usepackage{url}
\usepackage{color}

\usepackage{graphicx}
\usepackage[separate-uncertainty=true,multi-part-units=single,group-separator={,}]{siunitx}
\DeclareSIUnit[per-mode=symbol,per-symbol=p]{\ueV}{\micro\electronvolt}
\DeclareSIUnit[per-mode=symbol,per-symbol=p]{\um}{\micro\meter}

\usepackage{xspace}

\newcommand{\expv}[1]{\langle #1 \rangle}
\newcommand{\bs}[1]{\boldsymbol{#1}}

\newcommand{\Pth}{P_{\mathrm{th}}}

\begin{document}
\title{Ultralong temporal coherence in optically trapped exciton-polariton condensates}

\author{K. Orfanakis}
\affiliation{SUPA, School of Physics and Astronomy, University of St Andrews, St Andrews, KY16 9SS, United Kingdom}

\author{A.F. Tzortzakakis}
\affiliation{Institute of Electronic Structure and Laser, Foundation for Research and Technology – Hellas, GR-70013 Heraklion, Crete, Greece}

\author{D. Petrosyan}
\affiliation{Institute of Electronic Structure and Laser, Foundation for Research and Technology – Hellas, GR-70013 Heraklion, Crete, Greece}
\affiliation{A. Alikhanian National Science Laboratory, 0036 Yerevan, Armenia}

\author{P.G. Savvidis}
\affiliation{Institute of Electronic Structure and Laser, Foundation for Research and Technology – Hellas, GR-70013 Heraklion, Crete, Greece}
\affiliation{ITMO University, St. Petersburg 197101, Russia}
\affiliation{Westlake University, 18 Shilongshan Rd, Hangzhou 310024, Zhejiang, China} 
\affiliation{Westlake Institute for Advanced Study, 18 Shilongshan Rd, Hangzhou 310024, Zhejiang, China}

\author{H. Ohadi}
\email{ho35@st-andrews.ac.uk}
\affiliation{SUPA, School of Physics and Astronomy, University of St Andrews, St Andrews, KY16 9SS, United Kingdom}

\date{\today} 

\begin{abstract}
We investigate an optically trapped exciton-polariton condensate and observe temporal coherence beyond 1~ns duration. Due to the reduction of the spatial overlap with the thermal reservoir of excitons, the coherence time of the trapped condensate is more than an order of magnitude longer than that of an untrapped condensate. This ultralong coherence enables high precision spectroscopy of the trapped condensate, and we observe periodic beats of the field correlation function due to a fine energy splitting of two polarization modes of the condensate. Our results are important for realizing polariton simulators with spinor condensates in lattice potentials.
\end{abstract}
\maketitle

\section{Introduction}
The collectively enhanced accumulation of bosons in a single quantum state results in a coherent matter-wave known as a Bose-Einstein condensate (BEC). Coherence is a collection of correlations between the macroscopic multiparticle wavefunction of the condensate, and is a fundamental property of BECs. These correlations, which can be extended up to arbitrary orders~\cite{glauber_quantum_1963} can provide insight into the scattering processes in the condensate. First- and second-order correlation functions are the most studied quantities, quantifying phase and amplitude correlations. Phase coherence of atomic BECs has been studied both in space~\cite{andrews_observation_1997} and time~\cite{kohl_measuring_2001}. The advent of condensation in other bosonic platforms has enabled studying coherence in driven-dissipative systems such as semiconductor microcavity  exciton-polaritons (polaritons)~\cite{kasprzak_boseeinstein_2006, deng_condensation_2002, balili_bose-einstein_2007, deng_spatial_2007, love_intrinsic_2008, ohadi_spontaneous_2012, trichet_long-range_2013, fischer_spatial_2014, kim_coherent_2016, rozas_temperature_2018, rozas_determination_2019, klaas_evolution_2018} and photons~\cite{klaers_boseeinstein_2011, marelic_spatiotemporal_2016, schmitt_spontaneous_2016, damm_first-order_2017}. For polaritons in particular, early studies revealed an exponential decay of the first-order temporal correlation function with a coherence time of up to $\sim \SI{10} {ps}$~\cite{kasprzak_boseeinstein_2006, balili_bose-einstein_2007}. The reduction of intensity noise revealed a Gaussian decay with an improved coherence time of $\sim \SI{150}{ps}$ in single-spot excitation of polariton condensates~\cite{love_intrinsic_2008}. Recently, the temporal decay in a polariton laser with shot-noise-limited intensity stability displayed a transition from exponential to Gaussian with increasing condensation population, attributed to strong interactions within the condensate~\cite{kim_coherent_2016}.

Here, we measure the first order correlation function $g^{(1)}(t)$ in an optically trapped polariton condensate and observe temporal coherence beyond $\SI{1}{\ns}$, which is $\sim$ 20 times longer than that of an untrapped condensate. Unlike the previous works in untrapped condensates \cite{love_intrinsic_2008,kim_coherent_2016}, the theoretical fits of our data supports exponentially decaying correlations in the trapped condensate. Furthermore, we observe periodic oscillations of $g^{(1)}(t)$ due to the beating of two weakly-coupled polariton modes \cite{Kammann_SpinHall_2012}. The extraordinary long coherence time of the trapped polariton condensate allows observation of energy splittings as small as $\SI{16}{\ueV}$, which is 5 orders of magnitude smaller than the energy (chemical potential 1.54~eV) of the condensate. Our result thus enable a high resolution spectroscopy of polariton condensates, which can be applied to  precision tuning of the  condensates energies in optical lattices \cite{ohadi_spin_2017,OhadiSynchronizationcrossoverpolariton2018} and polariton simulators \cite{amo_exciton-polaritons_2016}.  

\section{Experiment}
\begin{figure}[t]
  \includegraphics[width=1\linewidth]{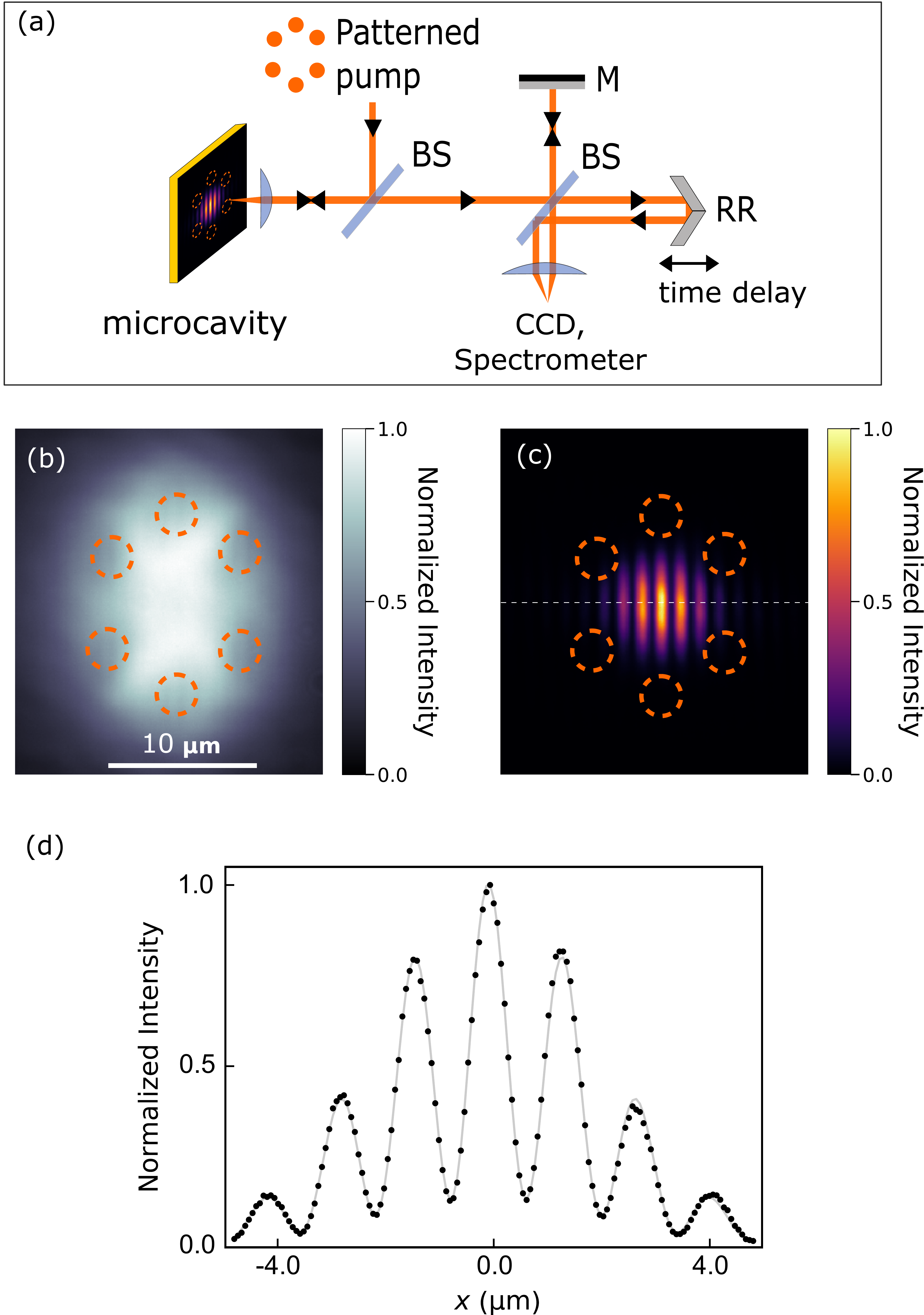}
  \caption{(a) The interferometer setup, with an adjustable time delay. BS: beam splitter, M: mirror, RR: retroreflector. Real space interferograms (b) below ($\sim 0.75\Pth$) and (c) above threshold ($\sim 2.50\Pth$). Pump spots of diameter $\SI{2.5}{\um}$ are marked by dashed circles. (d) Horizontal line profile across the dotted line of the interference pattern in (c). The gray line is the theoretical fit.}
  \label{fig:Interferogram}
\end{figure}

The sample is a 5$\lambda$/2 GaAs microcavity where polariton condensation under non-resonant optical pumping was previously observed~\cite{tsotsis_lasing_2012, ohadi_spontaneous_2015}. We excite polaritons using a single-mode quasi-continuous wave diode laser system. The laser is a home-made master oscillator power amplifier  composed of an external cavity diode laser seeding a tapered amplifier \cite{kangara_design_2014} (see Appendix \ref{App:Exp} for details). It is blue
detuned by 100~meV from the first Bragg mode of the mirror stop-band. An acousto-optic modulator (AOM) is used for laser power modulation and generation of $\SI{60}{\us}$ long pulses. The pulse duration is much longer than the rise/fall time of the AOM (\SI{40}{\ns}), and is considerably longer than the condensate formation time ($\sim 100$~ps) so that the condensate can be treated as stationary. 
Since the excitation duration is much shorter than typical vibration timescales ($\approx$1~ms), there is no need for the active stabilization of the interferometer~\cite{lagoudakis_quantized_2008}. We optically trap polaritons by patterning the laser radiation on the microcavity \cite{balili_bose-einstein_2007} using a spatial light modulator. To form the trap, the
laser beam is shaped into a hexagonal pattern of diameter $\approx \SI{7.5}{\um}$, see Fig.~\ref{fig:Interferogram}. 
Each pump spot creates an optically inactive reservoir composed of a hot
electron-hole plasma and dark excitons that repel and diffuse to the middle of the trap. During the diffusion,
this ‘inactive’ reservoir relaxes in energy and forms an optically active reservoir of excitons which couple to cavity photons and form polaritons in the middle of the trap. When the polariton density exceeds the condensation threshold, a macroscopically coherent condensate in the ground state of the momentum-space ($k$-space) forms in the center of the trap~\cite{cristofolini_optical_2013,askitopoulos_polariton_2013} (see Appendix \ref{App:Exp}).  The small overlap between the condensate and the hot reservoir at the pump spots results in a narrower linewidth in optically confined condensates compared to their unconfined counterparts~\cite{askitopoulos_polariton_2013}. In the untrapped geometry, the pump consists of only one spot and the condensate forms on top of the reservoir.

To study temporal coherence, we utilize a Michelson interferometer in the mirror-retroreflector configuration~[see Fig.~\ref{fig:Interferogram}(a)]. By varying the position of the moving arm and superimposing the image from each arm, we form interferograms for various time delays. Below the condensation threshold, emission is incoherent and no interference fringes are observed [Fig.~\ref{fig:Interferogram}(b)]. Above threshold, a macroscopically coherent state appears in the center of the trap attested by the presence of interference fringes [Fig.~\ref{fig:Interferogram}(c)]. A line profile in the center of the interferogram [Fig.~\ref{fig:Interferogram}(d)] can be fit by a Gaussian function (the condensate mode profile) multiplied by a cosine to acquire the fringe contrast~\cite{kim_coherent_2016}. The fringe contrast is equal to the magnitude of the first-order correlation function $\vert g^{(1)}(t)\vert$, where $t$ is the time delay. Since the condensate mode is small, spatial dependence in the correlation function is negligible. To acquire the temporal dependence of the correlation function, we record the fringe contrast while changing the time delay. We average over 5 realizations for each time delay.

\begin{figure}[t]
  \includegraphics[width=1\linewidth]{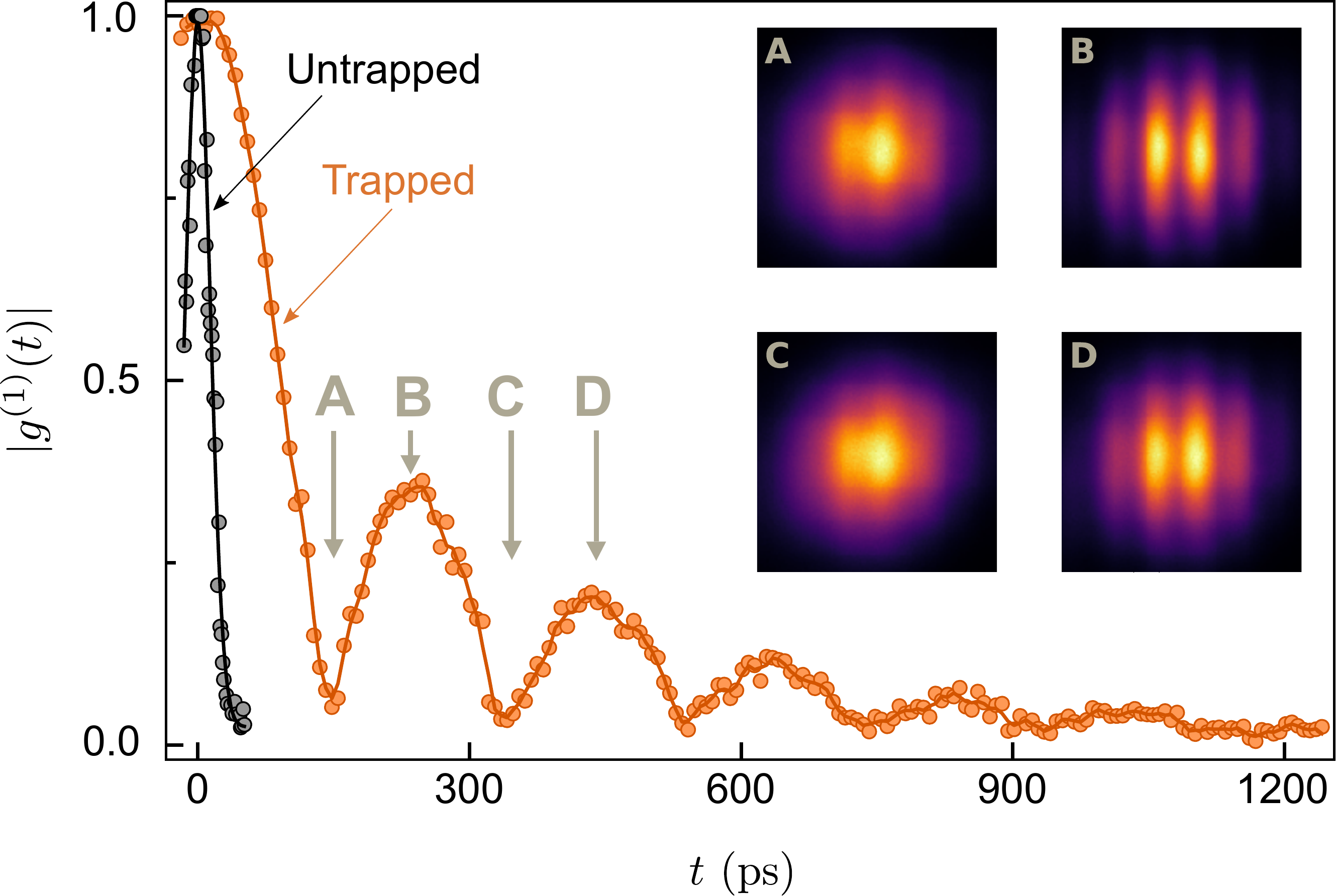}
  \caption{Temporal evolution of $\vert g^{(1)} \vert$ for an untrapped $P \approx 1.27\Pth$ (black circles) and a trapped condensate $P \approx 3.9\Pth$ (orange circles). Solid lines are guides to the eye. Inset: interferograms for four time delays (150, 250, 340 and $\SI{435} {ps}$), denoted as A, B, C and D.}
  \label{fig:Comparison}
\end{figure}

The effect of optical confinement on the polariton coherence is revealed by
comparing the decay of temporal coherence in the trapped and untrapped condensates
(Fig.~\ref{fig:Comparison}). The latter is characterized by a rapid decay of correlation function $g^{(1)}(t)$ (black curve in Fig.~\ref{fig:Comparison}) with a decay time
 ($1/e$ of fringe contrast) $\tau_{\mathrm{G}}\approx 20$~ps (see Fig.~\ref{fig:SI5} of Appendix \ref{App:Exp}).
The temporal correlation function of the trapped polariton condensate (orange curve in Fig.~\ref{fig:Comparison}), 
however, does not exhibit a monotonous decay as previously reported for untrapped 
condensates~\cite{kasprzak_boseeinstein_2006, balili_bose-einstein_2007, love_intrinsic_2008, kim_coherent_2016}. 
Instead, it exhibits a periodic oscillatory decay with a period of $\sim\SI{200} {ps}$. The contrast initially 
decays rapidly to nearly zero after $\SI{150} {ps}$, but then it recovers to 0.35 after 
$\sim\SI{100} {ps}$. We can observe a total of five peaks of $g^{(1)}(t)$ for up to $t = 1$~ns, approaching 
the exciton lifetime~\cite{hwang_lifetimes_1973, feldmann_linewidth_1987}. The coherence time for
the trapped condensate $\tau_{\mathrm{ph}}$ is more than an order of magnitude longer than that for 
the untrapped condensate (see below), which is due to the significant reduction of the spatial 
overlap  between  the  trapped  condensate  and  the  thermal  reservoir  of  excitons. 
The insets in Fig.~\ref{fig:Comparison} illustrate that the interferograms corresponding to the two first minima (A and C), 
have much lower fringe contrast than those corresponding to the first two maxima (B and D). 

\begin{figure*}[t]
  \includegraphics[width=0.9\linewidth]{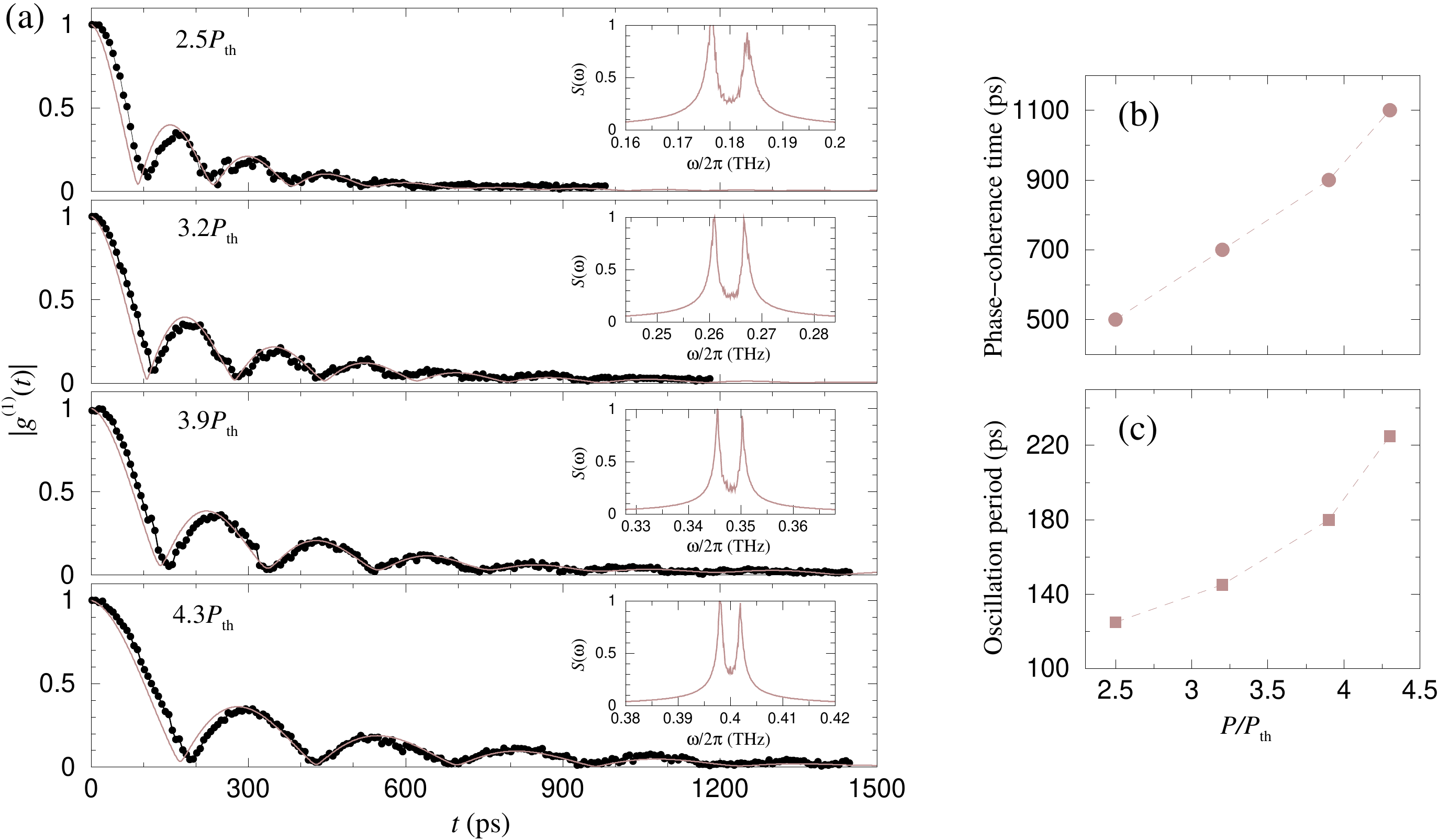}
  \caption{(a) Temporal decay of fringe contrast for four different pumping powers $P=2.5,3.2,3.9,4.3 P_{\mathrm{th}}$. 
    Circles represent experimental points, while solid lines are the correlation function $g^{(1)} (t)$ of Eq.~(\ref{eq:g1tau}) 
    as obtained from the numerical simulations of the polariton Eqs.~(\ref{eqs:psi12}) and (\ref{eq:Nexcit}) 
    averaged over long integration time. 
    Insets show the power spectrum $S(\omega)$ of the polariton field, as obtained via the Fourier transform of $g^{(1)} (t)$ 
    obtained from the numerical simulations, 
    illustrating the normal mode splitting by $\pm J$ around the central frequency $u(\bar{n}_1 +\bar{n}_2)$.  
    In the simulations, for each pumping power $P$ we used the phase-coherence time $\tau_{\mathrm{ph}} = 1/\kappa$ shown in (b),
    and the oscillation period $\pi/J$ shown in (c). }
  \label{fig:PDep}
\end{figure*}

The period of oscillations and the temporal decay of the correlation function depend on the excitation power. 
Increasing the pump power from $2.5\Pth$ to $4.3\Pth$ decreases both the frequency of oscillations $\Delta$ 
and the dephasing rate $\kappa = 1/\tau_{\mathrm{ph}}$, as shown in Fig.~\ref{fig:PDep}.

\section{Theory}

In the Michelson interferometry, the contrast of the real-space interferogram [Fig.~\ref{fig:Interferogram}(d)] 
is equal to the absolute value of the first-order field correlation function \cite{kim_coherent_2016}:
\begin{equation}
g^{(1)} (t) \equiv \frac{\expv{\bs{E}^*(r,t') \bs{E}(r,t+t')} }{ \sqrt{ \expv{|\bs{E}(r,t')|^2} \expv{ |\bs{E}(r,t+t')|^2 } }},
\label{eq:g1def}
\end{equation}
where $\expv{ \cdots }$ denotes the average over time $t'$.
The field $\bs{E}(r,t)$ is proportional to the exciton-polariton condensate wavefunction. 
We assume that there are two (nearly) energy-degenerate modes $\psi_1$ and $\psi_2$ of the polariton 
condensate with orthogonal circular polarizations $\bm{}\bm{\varepsilon}_1$ and $\bm{\varepsilon}_2$ 
($\bm{\varepsilon}_1 \cdot \bm{\varepsilon}^*_2 =0$). The total field is then
$\bs{E}(r,t) \propto \bm{\varepsilon}_1 \psi_1 (t) + \bm{\varepsilon}_2 \psi_2 (t)$. 
Substituting this into Eq.~(\ref{eq:g1def}), we have
\begin{widetext}
\begin{equation}
g^{(1)} (t) 
= \frac{\expv{ \psi_1 (t+t') \psi_1^* (t') +  \psi_2 (t+t') \psi_2^* (t') }}
{\sqrt{ \expv{ |\psi_1(t')|^2 + |\psi_2(t')|^2 }
\expv{ |\psi_1(t+t')|^2 + |\psi_2(t+t')|^2 }}} . 
\label{eq:g1tau}
\end{equation}

We use the standard approach \cite{Wouters2008} to describe the two polariton modes via the differential equations
\begin{subequations}
\label{eqs:psi12}
\begin{eqnarray}
\partial_t \psi_1 &=& \frac{1}{2}(R_1 N_1 - \gamma_1) \psi_1 - i [ \epsilon_1(t) + u_{11} |\psi_1|^2 + u_{12} |\psi_2|^2 ] \psi_1 
+ i J \psi_2  , \\
\partial_t \psi_2 &=& \frac{1}{2}(R_2 N_2 - \gamma_2) \psi_2 - i [ \epsilon_2(t) + u_{21} |\psi_1|^2 + u_{22} |\psi_2|^2 ] \psi_2 
+ i J^* \psi_1  ,
\end{eqnarray}
\end{subequations}
\end{widetext}
where $R_{j} N_{j}$ is the pumping rate of $j$th polariton from the reservoir of $N_{j}$ excitons, 
$\gamma_{j}$ is the polariton decay rate, $\epsilon_{j}$ are the single-particle energies, 
$u_{ij}$ are the non-linear self ($i=j$) and cross ($i\neq j$) interactions rates, and 
$J$ is the Josephson coupling between the two modes due to the spin-orbit interaction \cite{Kammann_SpinHall_2012}. 
The above equations are supplemented by the reservoir equations
\begin{equation}
\partial_t N_j = P_j - \Gamma_j N_j - R_j N_j |\psi_j|^2  , \label{eq:Nexcit} \\
\end{equation}
where $P_j$ is the thermal exciton pumping rate, $\Gamma_j$ is the decay (recombination) rate, and $R_j$ is the scattering rate
of the excitons into the BEC of polaritons. In the steady state, $\partial_t N_j = 0$ we have 
$N_j = \frac{P_j}{\Gamma_j + R_j |\psi_j|^2}$. Upon substitution into the polariton equations and demanding that
$R_j N_j - \gamma_j \geq 0$ we obtain the threshold pumping rate $P_j \geq \gamma_j  \Gamma_j /R_j$ for non-zero polariton
intensity $|\psi_j|^2 >0$. Above threshold, neglecting the coupling between the two polaritons, the average polariton 
intensity in the steady state is then $\bar{n}_j \equiv \expv{ |\psi_j|^2} \simeq \frac{P_j}{\gamma_j} - \frac{\Gamma_j}{R_j}$, 
while the exciton number is $N_j \simeq \frac{\gamma_j}{R_j}$. 

We assume that each polariton mode is subject to phase fluctuations with the rate $\kappa = 1/\tau_{\mathrm{ph}}$,
which would correspond to exponential decay of polariton coherence. 
We model these phase fluctuations by a Wiener process: we take $\epsilon_{j}(t)$ in Eqs.~(\ref{eqs:psi12}) 
to be Gaussian stochastic variables with the mean $\expv{\epsilon_{1,2}} =  0$ 
and variance $\sigma^2_{\epsilon} = 2\kappa/\delta t$, where $\delta t$ is the time increment 
in the simulations of the system dynamics. 
We simulate the polariton Eqs.~(\ref{eqs:psi12}) numerically, staring with small random seed 
amplitudes $\psi_{1,2} \neq 0$ at some time $t_0$\footnote{Simulation parameters are (all in units of 1/ps): 
$\gamma_{1,2} = 1$, $\Gamma_{1,2} = 0.3$, $R_{1,2} = 0.001$ yielding $\Pth = \gamma \Gamma/R = 300$ and thus 
$P = 2.5, 3.2, 3.9, 4.3 \Pth$ leading to $\bar{n}_{1,2} \simeq 450, \, 660, \, 870, \, 990$ after a short transient. 
We set all the non-linear self and cross interaction coefficients the same, $u_{ij} \equiv u = 2\times 10^{-4}$,
which shifts the mean energy of the polariton condensate but does not change the energy difference 
between the two polariton modes. The remaining fitting parameters are $J = \Delta/2$ and $\kappa$.}. 
The initial values of the seed are unimportant, as after a short transient of a few tens of ps duration 
the polariton populations attain close to the steady-state values $\bar{n} \simeq \frac{P}{\gamma} - \frac{\Gamma}{R}$ 
determined by the exciton pumping rate (provided $P \geq \Pth = \gamma \Gamma/R$).
The correlation functions (\ref{eq:g1tau}) are then obtained upon long-time averaging over the system dynamics.
To fit the experimental data in Fig.~\ref{fig:PDep},
for each pumping rate $P_j$ we set $J \simeq \Delta/2$ and choose appropriate dephasing rate $\kappa$. 
The periodic oscillations of the field correlation function then corresponds to the beating   
of the two eigenmodes split by $\pm J$ and broadened by $\kappa$. This is illustrated by the power spectrum 
of polariton field $S(\omega)$ given by the Fourier transform of $g^{(1)} (t)$ [see the insets in Fig.~\ref{fig:PDep}(a)].
The spectrum $S(\omega)$ is centered at frequency $u(\bar{n}_1 +\bar{n}_2)$ and split by $\pm J$.

The observed energy splitting $\Delta \simeq 2J$ of the two polariton eigenmodes decreases with increasing the pumping power 
and thereby the polariton intensity $\bar{n}$. Our simulations of the Gross-Pitaevskii equations for the 
polariton BEC indicate that with increasing the pumping power the lateral size of the polariton BEC in real-space
is progressively increased, while its in-plane $k$-space distribution is correspondingly reduced. 
This explains the reduction of the Josephson coupling between the two polariton components due 
to the spin-orbit interaction between the two bare polariton modes \cite{Kammann_SpinHall_2012}. 
We also note that according to the arguments in  \cite{kim_coherent_2016}, the exponential decay 
of polariton coherence has a rate $\kappa = 1/\tau_{\mathrm{ph}} \sim \gamma/2\bar{n}$, with $\gamma$ 
being the decay rate of the polariton intensity; hence, the polariton coherence time is proportional 
to its intensity, $\tau_{\mathrm{ph}} \propto \bar{n}$,  and thereby the pumping power (stronger pumping 
-- longer coherence time), consistent with our numerical simulations.

\section{Conclusions} 

To summarize, we have demonstrated that optically trapping an exciton-polariton condensate increases the coherence time by more than an order of magnitude compared to that of an untrapped condensate. In the untrapped case, the condensate is formed on top of a sea of hot reservoir excitons that directly interacts with the condensate, and causes strong decoherence. In the optically trapped condensate, however, this hot reservoir is mostly spatially decoupled from the condensate. This critical difference permits observations of ultralong coherence times in the trapped condensate.  The highly prolonged coherence of the trapped polariton condensate allowed us to observe temporal beating of the first-order correlation function of the emitted field resulting from the fine structure in the condensate energy spectrum. This amounts to demonstration of measuring the energy splitting of the two interacting polarization modes of the condensate with unprecedented precision. Our results can thus be important for the characterization and control of spinor condensates in lattice potentials for realizing analog and digital polariton quantum simulators \cite{amo_exciton-polaritons_2016, lagoudakis_polariton_2017, berloff_realizing_2017, kalinin_networks_2018, kalozoumis_coherent_2020}.

\begin{acknowledgments}
The authors acknowledge fruitful discussions with Anton Nalitov, Alexey Kavokin, Michal Matuszewski and Jeremy Baumberg. This work was supported by Grant No.
EPSRC EP/S014403/1. K.O. acknowledges EPSRC for PhD studentship support through
grant no. EP/L015110/1. P.S. acknowledges support by Westlake University (Project No. 041020100118), Program 2018R01002 supported by Leading Innovative and Entrepreneur Team Introduction Program of Zhejiang, and bilateral Greece-Russia Polisimulator project co-financed by Greece and the EU Regional Development Fund.
\end{acknowledgments}

\appendix

\section{Experimental methods}
\label{App:Exp}

\subsection{Sample}

The sample constitutes a high-Q, $5\lambda/2$, GaAs-based microcavity with a top (bottom) DBR mirror comprising 32 (35) alternating layers of AlAs/Al$_{0.15}$Ga$_{0.85}$As. Four sets of three 10 nm Al$_{0.3}$Ga$_{0.7}$As/GaAs quantum wells are placed at the antinodes of the electromagnetic field inside the cavity. Measurements with different detuning are possible due to a wedge in the sample thickness which permits continuous tuning of the cavity mode with respect to the exciton mode. Polariton condensation under non-resonant optical pumping in this sample has been previously observed \cite{ohadi_spontaneous_2015,tsotsis_lasing_2012}.

\subsection{Experimental setup} \label{ES_overview}

\begin{figure}[t]
  \centering
  \includegraphics[width=1\linewidth]{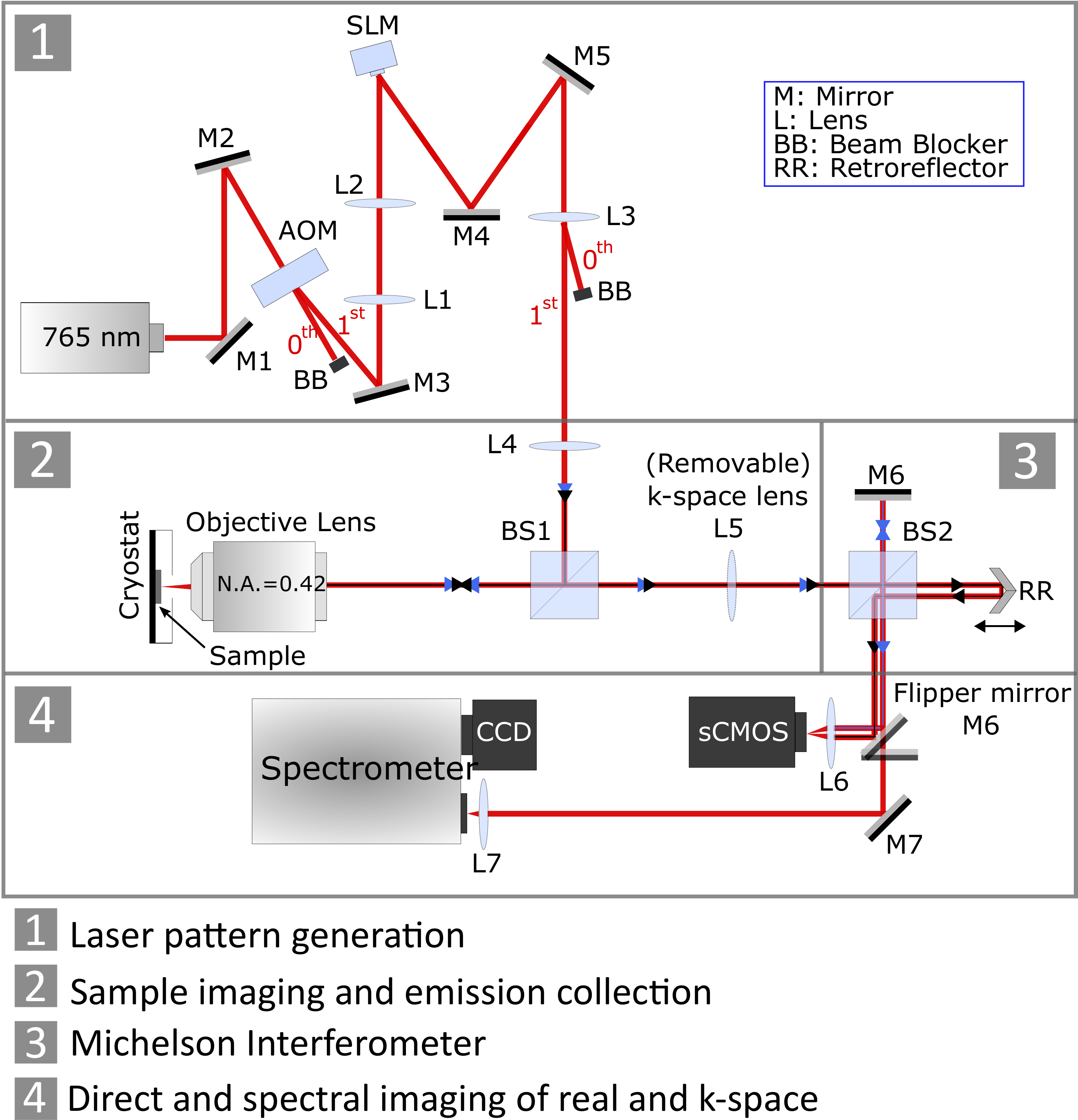}
  \caption{Experimental setup for real- and $k$-space analysis of polariton condensates. The focal length of each lens is: $f_{L1} = 50~$mm, $f_{L2} = 300~$mm, $f_{L3} = 400~$mm, $f_{L4} = 200~$mm, $f_{L5} \mathrm{(k-space)} = 300~$mm, $f_{L6} = 700~$mm, and $f_{L7} = 200~$mm. Distances are not drawn to scale. For the AOM and SLM, $0^{\mathrm{th}}$ and $1^{\mathrm{st}}$ correspond to the zeroth and first order transmission and reflection beam, respectively.}
  \label{fig:SI1}
\end{figure}

The experimental setup is schematically depicted in Fig.~\ref{fig:SI1} and consists of 4 main parts:
\begin{enumerate}
  \item
  \textbf{Laser pattern generation.} A tapered amplifier laser system emitting at 765 nm provides the pump laser beam (red) which is then directed into an acousto-optic modulator (AOM) for amplitude modulation. The final key-component is a spatial light modulator (SLM), whose function is generating the optical trapping potentials.
  \item
  \textbf{Sample imaging and emission collection.} A telescope formed by a spherical lens (L4) and a microscope objective is then used for scaling the image down to micrometer sizes and projecting it onto the microcavity sample mounted inside a cryostat. Polariton emission is then collected from the objective in a backscattering geometry and directed towards the interferometer. A removable lens (L5) is included for $k$-space (momentum-space) measurements.
  \item
  \textbf{Phase measurements with an interferometer.} Temporal coherence measurements are performed by implementing a Michelson interferometer in the mirror-retroreflector configuration. The retroreflector is placed on a linear translation stage equipped with an integrated electron controller, thus allowing the automated movement of the retroreflector.
  \item
  \textbf{Direct and spectral imaging of real and $k$-space.} The last part of the setup is for imaging either 2D real-space onto a scientific CMOS (sCMOS) camera or k-space onto the CCD camera of a spectrometer.
\end{enumerate}

\subsection{External cavity diode laser and tapered amplifier laser system}

\begin{figure}[t]
  \centering
  \includegraphics[width=1\linewidth]{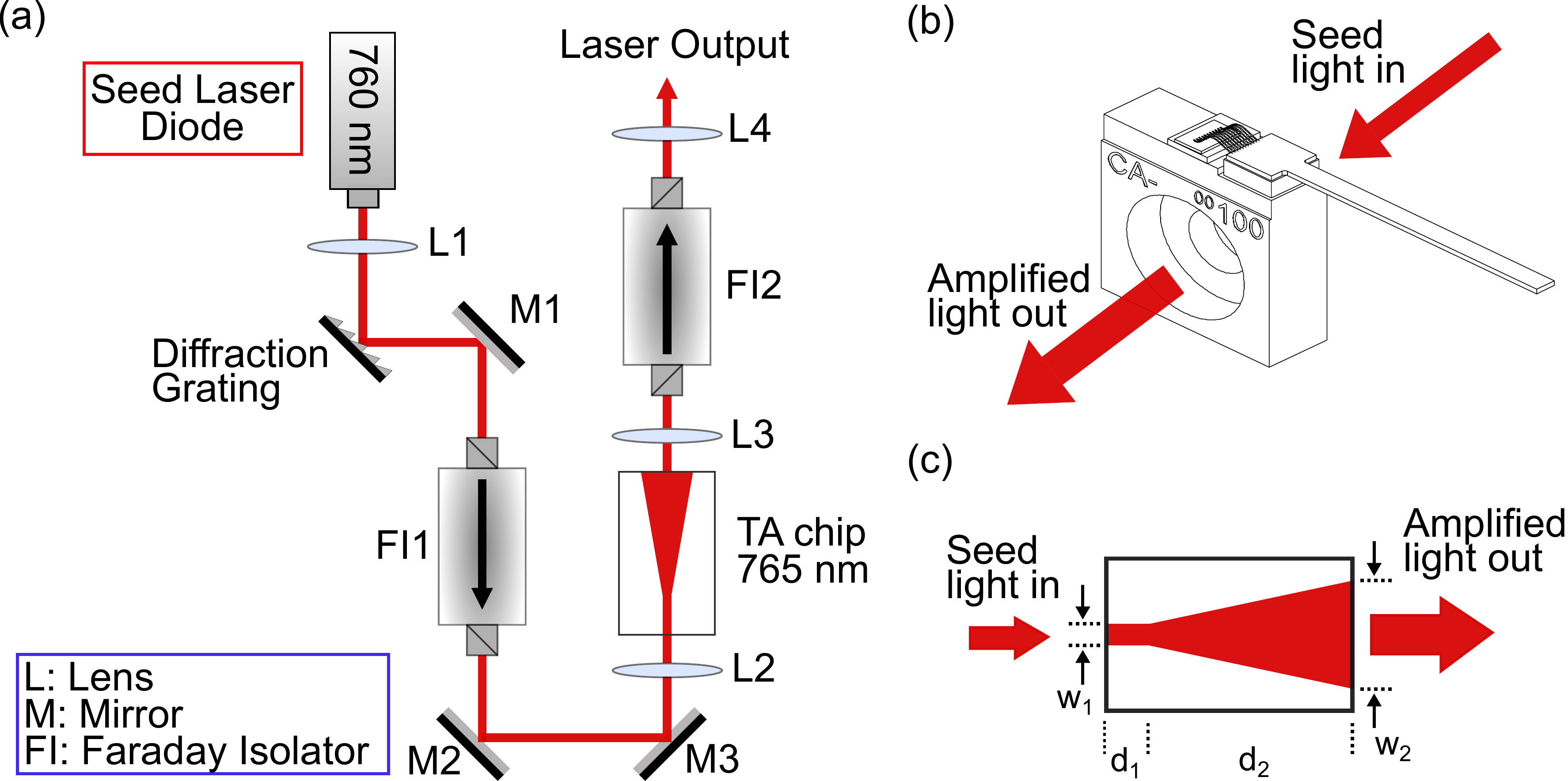}
  \caption{(a) The TA laser system. The focal length of the three aspheric lenses, L1-3, is 4.51 mm. 
    For the cylindrical, L4, $f = 50~$mm. The diffraction grating is a reflective holographic grating with 1800 grooves/mm.
    (b) Schematic drawing of the TA chip.
    (c) Schematic drawing of the top-view of the TA chip where the straight, index-guided section of the chip of length $d_{1}$ and width $w_{1}$, and the tapered, gain-guided section of length $d_{2}$ and output width $w_{2}$ are depicted. Typical values are: $d_{1} \approx 0.5~$mm, $d_{2} \approx 1.5 - 3~$mm, $w_{1} \approx 1-3$~$\mathrm{\upmu m}$ and $w_{2} \approx 200$~$\mathrm{\upmu m}$ \cite{kangara_design_2014}.}
\label{fig:SI2}
\end{figure}

An integral component of this laser system is the external-cavity diode laser (ECDL)~\cite{kangara_design_2014, arnold_simple_1998, nyman_tapered-amplified_2006}. In this setup, a laser diode emitting at 760 nm is initially mounted in a tube equipped with an aspheric collimation lens with a focal length $f = 4.51~$mm [denoted as L1 in Fig.~\ref{fig:SI2}(a)]. The external cavity is built in a modified Littrow configuration in which the collimated beam is reflected by a grating with the first diffraction mode retroreflected back into the diode and the zeroth order used as output~\cite{maddaloni_laser-based_2013}. Optical feedback with a narrow-linewidth laser output is thus established between the end facet of the laser diode and the diffraction grating, which form an external cavity. Moreover, only a narrow range of wavelengths reflects back to the diode for amplification, because of the wavelength-selective reflectivity of the grating. As the wavelength is tuned, however, the direction of the output beam changes. For this reason, a mirror (M1) is placed on the same stage as the grating to compensate for grating adjustments. Besides tunability and narrow-linewidth, an ECDL constitutes a low-cost and compact laser device offering high stability~\cite{duarte_tunable_2017}.

The requirement for single-mode operation requires the transverse dimension of the laser diode to be of the order of the optical wavelength, which limits the output of the ECDL ($\sim$100 mW). In our case, this restriction was circumvented by implementing a tapered amplifier (TA) system for amplifying the low output of the seed ECDL, while simultaneously preserving its spectral properties (wavelength, linewidth etc.) \cite{kangara_design_2014, walpole_semiconductor_1996}. In the TA chip light from the seed is initially inserted from the narrow aperture into a short, straight, index-guided section [Fig.~\ref{fig:SI2}(b) and (c)]. The narrow transverse dimension of this region permits the excitation of only the fundamental transverse mode, thus ensuring high beam quality. The light radiation is then guided into a longer, tapered, gain-guided region for optical amplification. This region is typically made of a III-V semiconductor (AlGaAs in our case). While in operation, carrier population inversion is established through uniform electrical pumping of the tapered region. Consequently, optical amplification of the propagating beam is achieved through the mechanism of stimulated emission from electron-hole recombination, just as in conventional lasing. One basic requirement is matching the wavelength of the tapered amplifier to the emission of the seed laser, hence a TA chip with a center wavelength of 765 nm is used for amplifying the ECDL. The laser after the TA has a maximum output of ${\sim}2~$W.

The design of the TA chip is such that the propagating beam is freely diverging in the tapered plane until it is emitted through the much wider aperture at the end of the tapered section [Fig.~\ref{fig:SI2}(c)]. An aspheric lens ($f = 4.51~$mm) is placed on either side of the TA chip, the first (L2) for focusing the input and the other (L3) for collimating the vertical component of output beam. Apart from high divergence, the output beam is highly astigmatic. Therefore, an additional lens with a cylindrical shape (L4) is placed after L3 for collimating the horizontal component of the beam.

A small portion of the amplified radiation is reflected back from the front facet of the TA chip, which could disturb the single-mode operation of the seed laser. A Faraday isolator is placed between the TA and the ECDL in order to suppress this back reflected radiation. Furthermore, optical backreflections can potentially cause permanent damage to TA chips, therefore another Faraday isolator (FI2) is placed after the TA chip to prevent this.

Finally, for stable operation the temperature is regulated. Specifically, maintaining the temperature of the seed laser diode constant ensures power and frequency stability. Additionally, the TA chip needs to be actively cooled since ${\sim} 10$~W of power is electrically pumped into the chip ($I = 3.8~$A, $V = 2.9~$V) but only ${\sim} 2$~W of optical power is extracted. For this purpose, a thermoelectric cooling device is placed below the mounts for both the diode and the TA chip.

Our laser system provides several advantages over other lasers commonly used within the polariton community, e.g. Ti:Sapphire \cite{ohadi_spontaneous_2015}. Most importantly, 1ow intensity noise operation makes it an ideal laser system for interferometric measurements. Because random fluctuations in the laser power are minimized, the dephasing induced by our apparatus is minimized as well. This in fact has been demonstrated using a solid-state laser diode pump \cite{love_intrinsic_2008}. Moreover, the ECDL+TA laser is a compact device with both low construction cost ($ {\sim} \pounds 7,500 $) and low maintenance. However, its tuning range ($ {\sim}10$~nm) is considerably less to that offered by Ti:Sapphire lasers (${\sim} 100~$nm).

\subsection{Power dependence and $k$-space emission}

\begin{figure}[t]
  \centering
  \includegraphics[width=1\linewidth]{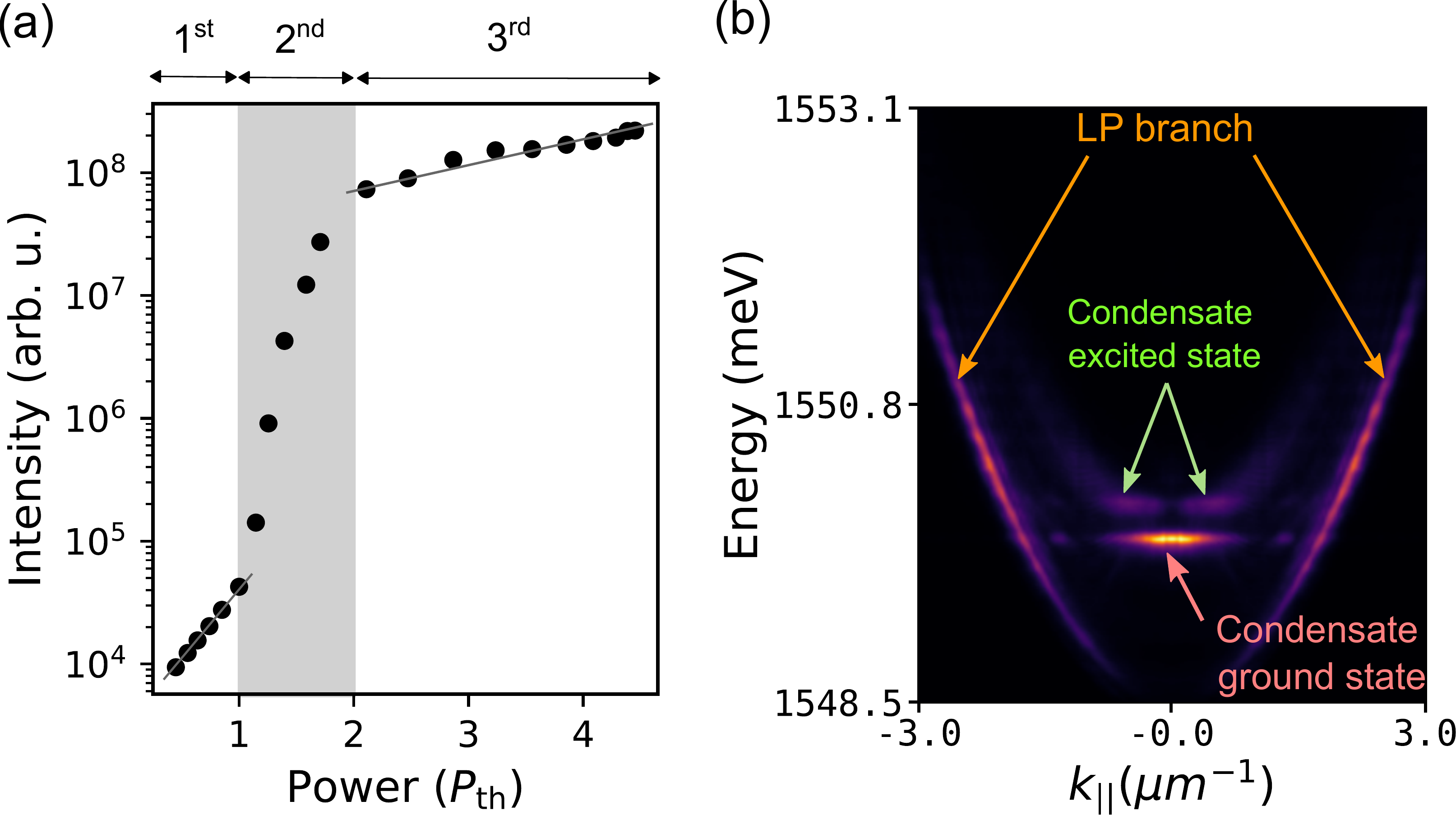}
  \caption{(a) Power dependence of the cavity PL emission. The grey lines are used as guide to the eye, illustrating where the emission intensity grows linearly with the pump, while the grey-shaded area denotes the non-linear response of the PL emission associated with the onset of condensation.
        (b) Dispersion curve for a pump power ${\sim} \Pth$ illustrating all light-emitting entities in the microcavity sample.}
  \label{fig:SI4}
\end{figure}

The dependence of photoluminescence on pump power exhibits three distinct regimes [see Fig.~\ref{fig:SI4}(a)]. The low-power regime corresponds to conventional polariton decay and increases linearly with the excitation power. The second regime is the onset of condensation at a critical power threshold, $\Pth$. This is accompanied by a nonlinear increase of the emission up until ${\sim} 2\Pth$ and then is followed by a sublinear third regime. An interesting feature is the sublinear behavior of the third regime, contrary to what has been observed for untrapped polariton condensates~\cite{tsotsis_lasing_2012}.

Figure~\ref{fig:SI4}(b) shows the $E(k_{\parallel})$ dispersion curve for a pump power near threshold, where $k_{\parallel}$ is the in-plane wavevector. Most of the emission comes from the condensed polariton population, which is blueshifted by ${\sim} 1.2~$meV relative to the bottom of the LP branch due to polariton self-interactions within the condensate mode and polariton interactions with the exciton reservoir \cite{bajoni_polariton_2008-1}. Uncondensed higher-$k_{\parallel}$ polaritons also contribute to the emission from the cavity. Near threshold we can observe the first excited condensate mode at energies ${\sim} 0.4~$meV above the principal one, although much weaker. We note that the energy splitting between the ground state and excited state is two orders of magnitude larger than the energy splitting responsible for our coherence oscillations.

\subsection{Untrapped condensate coherence time}

\begin{figure}[t]
  \centering
  \includegraphics[width=0.50\linewidth]{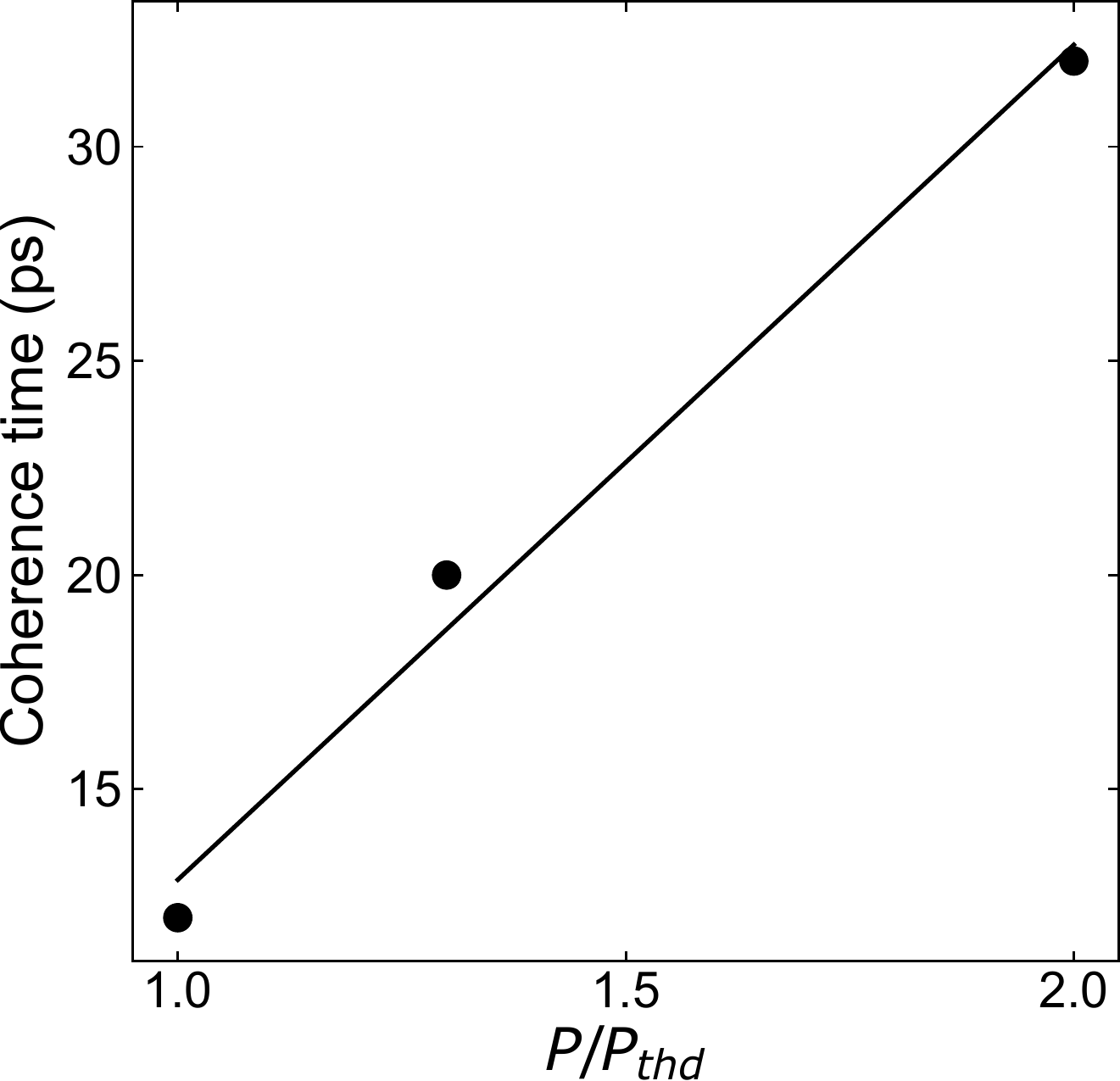}
  \caption{Power dependence of coherence time $\tau_{\mathrm{G}}$ for the untrapped condensate. }
  \label{fig:SI5}
\end{figure}

Power dependence of coherence time was also measured for untrapped condensates, see Fig.~\ref{fig:SI5}. Specifically, coherence time is equal to 12 ps just above threshold and almost triples for a pumping power ${\sim} 2\Pth$. This coherence time is more than an order of magnitude smaller than that for trapped condensates at all powers.

\end{document}